\documentclass[12pt,preprint]{aastex}

\def\cm{{\,\rm cm}}
\def\kev{{\,\rm keV}}
\def\kpc{{\,\rm kpc}}
\def\kms{{\rm km\,s^{-1}}}

\def\ergs{{\rm erg\,s^{-1}}}

\def\Lx{L_{\rm X}}

\def\L6.3{L_{\rm 6.3cm}}

\def\NH{N_{\rm H}}

\def\keV{\,{\rm keV}}

\def\date         {\ifcase\month \message{zero} \or
                    January \or February \or March \or April \or May \or June
                    \or July \or
                    August \or September \or October \or November \or
                    December \fi
                    \space\number\day, \number\year}
\def\Halpha{{\rm H}\alpha}
\def\Feka{{\rm Fe\, K}\alpha}
\def\Mgka{{\rm Mg\, K}\alpha}
\def\Sika{{\rm Si\, K}\alpha}

\def\pcmcu{\hbox{$\cm^{-3}$}}

\begin{document}
\title{CHANDRA OBSERVATIONS OF ULIRGS: EXTENDED HOT GASEOUS HALOS IN MERGING
GALAXIES}

\author{Z.Y. Huo$^{1}$, X.Y. Xia$^{1,2}$, S.J. Xue$^{1}$, S. Mao$^{3}$,
and Z.G. Deng$^{1,4}$}

\altaffiltext{1}{National Astronomical Observatories,
                Chinese Academy of Sciences, A20 Datun Road, 100012 Beijing,
                China}
\altaffiltext{2}{Dept. of Physics, Tianjin Normal University,
        300074 Tianjin, China}
\altaffiltext{3}{Univ. of Manchester, Jodrell Bank Observatory,
          Macclesfield, Cheshire SK11 9DL, UK}
\altaffiltext{4}{Dept. of Physics, Graduate School,
        Chinese Academy of Sciences, 100039 Beijing, China}

\received{\date}
\accepted{ }

\begin{abstract}
We study the properties of hot gaseous halos in 10 nearby ultraluminous
IRAS galaxies observed with the ACIS instrument on board {\it Chandra}.
For all sample galaxies, diffuse soft X-ray emissions are found
within $\sim$ 10\,\kpc\ of the central region; their spectra are well fitted by
a MEKAL model plus emission lines from $\alpha$-elements and other ions. The
temperature of the hot gas is about 0.7\,keV and metallicity is about
$1 Z_\odot$.
Outside the central region, extended hot gaseous halos are found for nine
out of the ten ULIRGs. Most spectra of these extended halos can be fitted with
a MEKAL model with a temperature of about 0.6\,keV and a low metallicity
($Z \sim 0.1 Z_\odot$).
We discuss the implications of our results on the origin of X-ray halos
in elliptical galaxies and the feedback processes associated with starbursts.
\end{abstract}


\section{INTRODUCTION}

It is now widely accepted that ultraluminous infrared galaxies (hereafter
ULIRGs) are associated with galaxy interactions/mergers (see Sanders \&
Mirabel 1996 for a review). They belong to a population of objects that
are en route to form elliptical galaxies through massive starbursts and
active galactic nucleus (AGN) phases (e.g., Sanders et al. 1988;
Melnick \& Mirabel 1990). Although the nature of ULIRGs is understood in
general terms, it is still unclear whether all these objects originate from
the interactions/mergers of a pair of gas-rich spirals or some may form from
galaxy groups. The latter would imply that some ULIRGs will become giant
elliptical galaxies through a bright QSO phase with galaxy groups as their
progenitors.

There is observational evidence to support the scenario of elliptical galaxy
formation through multiple merging in both clusters of galaxies and in the
field. For example,  high resolution infrared and optical observation
identify more than two luminous nuclei in some merging galaxies (e.g., Seigar,
Lynam \& Chorney 2003; Laine et al. 2003; Borne et al. 2000; Cui et al.
2001 and references therein). If these are genuine galactic nuclei, the group
connection is then clearly established. However, the issue is clouded by the
fact that it is difficult to separate galaxy nuclei from massive
super-star-clusters that also appear frequently in massive starburst galaxies.

X-ray observations can provide an independent way to distinguish the
progenitors of mergers because if the progenitor of a merger is a group of
galaxies, the X-ray halo of the group may survive quite a long time after
merging. Fossil groups have generated much recent interests. They are
dominated by central luminous giant elliptical galaxies and have luminous
spatially extended X-ray emissions. They are thought to be the end result
of merging within normal groups (Jones et al. 2003). If the extended
diffuse X-ray emissions of fossil groups and ULIRGs are similar, then this
would suggest an evolution link between ULIRGs and fossil groups (Borne et
al. 2000; Cui et al. 2001; Jones et al. 2003).

On the other hand, elliptical galaxies are known to have hot gaseous halos
(see Mathew \& Brighenti 2003 for a review). In the hierarchical structure
formation theory, elliptical galaxies are formed by galaxy merging, and so
the origin of such hot gaseous halos and the metal enrichment may be closely
related with the merger process. Luminous elliptical galaxies show very 
extended diffuse X-ray emissions, while the X-ray emissions for faint 
ellipticals are mainly contributed by discrete sources, such as X-ray 
binaries (O'Sullivan, Forbes \& Ponman 2001). Furthermore, the X-ray 
luminosities of bright ellipticals have large scatters (Matsushita 2001; 
O'Sullivan et al. 2001; Sansom, Hibbard \& Schweizer 2000). From 
{\it Chandra} observations, it is now clear that the spectra of low-mass 
X-ray binaries can be fit by either a thermal bremsstrahlung model with 
$kT \sim 7\keV$ or a power-law with $\Gamma\approx 1.6$ (e.g., Irwin, 
Sarazin \& Bregman 2002, Fabbiano et al. 2004). Consequently, the diffuse 
hot gas component of early type galaxies could be determined by 
multi-components spectral and spatial fitting analysis. Based on a careful 
spectral and spatial analysis for 39 X-ray luminous early type galaxies 
observed by {\it ROSAT} PSPC,
O'Sullivan, Ponman \& Collins (2003a) suggested that the X-ray halos
of elliptical galaxies are not formed by the infall and shock heating of
primordial gas in clusters and groups of galaxies, instead, they may be
produced by the stellar mass loss within galaxies. However, O'Sullivan et al.
(2003a) found that the metallicities of luminous early type galaxies cover
a broad range, namely from very low metallicity ($0.1Z_\odot$) to high
metallicity ($>1 Z_\odot$). It seems difficult to produce low metallicity
hot gas from stellar mass loss. The diverse properties of hot gaseous halos in
ellipticals imply that their progenitors and the origins of the hot gas may
not be unique.

The energy feedback and metal enrichment processes associated with
stellar mass loss, superwinds and AGNs are still not well understood in galaxy
formation. {\it Chandra} X-ray Observatory, with its sub-arcsecond resolution
and high-energy sensitivity, is an ideal instrument to probe these processes
in massive starburst galaxies such as ULIRGs. From {\it Chandra} observations
of three ULIRGs, NGC 6240, Mrk 273 and Arp 220 (Lira et al. 2002; Xia et al.
2002; McDowell et al. 2003), it appears that the diffuse soft X-ray emissions
in the inner $\sim$ 10 \kpc\ scale is spatially correlated with the $\Halpha$
filaments and arcs, similar to that found in nearby starburst/normal
galaxies (Strickland et al. 2003). However, the more extended (out to few
tens of \kpc) hot gaseous halo in Mrk 273 has no corresponding $\Halpha$
emission. Furthermore, the metallicity for this gas is much lower than that 
for the inner hot gas. This implies that the metals produced and ejected in 
central star formation regions have not yet enriched the gas in the outer 
region ($\ga 10\kpc$) after the system has almost finished merging.

{\it Chandra} and {\it XMM-Newton} have observed more than 15 luminous or
ultraluminous IRAS galaxies (Ptak et al. 2003; Franceschini et al. 2003).
Diffuse thermal emission components are detected in all the observed ULIRGs.
Furthermore, these observations also reveal very extended diffuse soft X-ray
emissions for NGC 6240, Mrk 231 and IRAS 19254-7545  (Lira et al. 2002;
Gallagher et al. 2002; Franceschini  et al. 2003), similar to that in
Mrk 273 (Xia et al. 2002). This sample, which includes ULIRGs at different
merging stages, provides a valuable database to understand the origin of hot
gas halos in elliptical galaxies as well as the chemical enrichment and energy
feedback processes. Analyses of this sample can also shed some light on
similar processes occurring in their high-redshift counterparts, such as the
SCUBA sources (e.g., Sanders 1999).

There are already two studies that present the statistical results based on
ULIRGs sample observed by {\it Chandra} (Ptak et al. 2003) and
{\it XMM-Newton} (Franceschini et al. 2003). However, these two previous
papers are mainly concerned about the relative contributions to the X-ray
emission by AGNs and starbursts. This study concentrates more on the
properties of diffuse hot gas in ULIRGs. The structure of the paper is as
follows. In \S 2, we describe the sample, observations and data reduction.
In \S 3, we present the images and surface brightness profiles for all sample
ULIRGs. In \S 4, we study the spatially-resolved spectral behaviors for the
extended soft X-ray emissions. Finally, in \S5, we summarize our results and
discuss the implications on the origin of the hot gas in elliptical galaxies
and feedback processes. Throughout this paper, we use a cosmology with a Hubble
constant $H_0=65\,\kms {\rm Mpc}^{-1}$, $\Omega_{\rm m}=0.3$ and
$\Omega_\lambda=0.7$; the values of $\Omega_{\rm m}$ and $\Omega_\lambda$ have
little influence on our results due to the low-redshift ($z < 0.045$) of our
sample ULIRGs.

\section{SAMPLE AND DATA REDUCTION}

In order to study the properties of diffuse hot gaseous halos in merging
galaxies and the associated energy feedback, metal enrichment processes in
massive starburst galaxies, we need a spatially resolved sample of merging
galaxies with sufficient number of detected photons. For this purpose, we
use the sample taken from the {\it Chandra} archive compiled by Ptak et al.
(2003) plus Arp 299 (Zezas, Ward \& Murray 2003). Although the far-infrared
luminosities of Arp 299 and NGC 6240 are slightly below the formal definition
of ULIRGs, they are merging galaxies with two separated nuclei and have
properties similar to ULIRGs in all wavebands. So for convenience we will
also loosely group them together with ULIRGs. The sample includes a total of
10 ULIRGs, of which 4 have a single nucleus and 6 have two nuclei. Seven of
these ULIRGs have redshift close to 0.04 while the remaining three have
redshift smaller than 0.025. In Table 1, we list the observation log for the
ULIRGs observed by {\it Chandra}. All the data except Mrk 273 are retrieved
from the {\it Chandra} archive. Arp 299 was observed by the ACIS-I detector
while all the other galaxies were observed with ACIS-S.

The data reduction is performed using {\it Chandra} Interactive Analysis of
Observations package (CIAO) v2.3 and the data are reprocessed using CALDB
2.21 from the level 1 event file. For observations taken in the VFAINT mode,
we run the tool {\itshape acis\_process\_events\/} to flag probable
background events using all the information of the pulse heights in a
$5 \times 5$ event island (as opposed to a $3 \times 3$ event island
recorded in the FAINT mode) to  distinguish good X-ray events and bad events
that are most likely associated with cosmic rays. We also apply the charge
transfer inefficiency correction when the temperature of the focal plane at
the time of the observation was $-120^\circ$\,C. This procedure allows us to
recover the original spectral resolution partially lost due to the charge
transfer inefficiency. The background light curve is examined and time of
background flaring is removed when the background count rate exceeds 1.2
times the mean value. The data are filtered to include only the standard
event grades 0, 2, 3, 4, 6. Point sources are detected using
{\it wavdetect\/}.

Diffuse X-ray emissions are detected for all sample sources. The data were
analyzed in full (0.3-8.0 keV), soft (0.3-2.0 keV) and hard (2.0-8.0 keV)
bandpasses, respectively. We will present the X-ray imaging and spectral
analyses in \S\,3 and 4.

\section{X-RAY IMAGES AND SURFACE BRIGHTNESS PROFILES}

We extracted the images in the soft (0.3-2.0\keV) and hard X-ray
(2.0-8.0\keV) bands and performed exposure correction according to the way 
adopted by Strickland et al. (2003). Each exposure-corrected 0.3-2.0 keV 
energy band image is constructed from the exposure-map-corrected images in 
the 0.3-0.6, 0.6-1.0, 1.0-1.5 and 1.5-2.0 keV energy bands while each 
exposure-corrected 2.0-8.0 keV energy band image is constructed from the 
exposure-map-corrected images in the 2.0-4.0, 4.0-6.0 and 6.0-8.0 keV energy 
bands. These exposure corrected 0.3-2.0 keV images were adaptively smoothed 
by using of the CIAO task {\it csmooth\/} at a $3\sigma$ significance level 
and a maximum smoothing scale about 10 pixels ($\approx$ 5$\arcsec$). 
Fig. Fig. 1 shows the adaptively smoothed soft (0.3-2.0\keV) 
X-ray images (left) and optical images overlaid by soft X-ray contours 
(right) for the sample ULIRGs. Most optical images are from {\it HST} 
archive\footnote{Based on observations made with the NASA/ESA Hubble Space 
Telescope, obtained from the data archive at the Space Telescope Institute. 
STScI is operated by the association of Universities for Research in 
Astronomy, Inc. under the NASA contract NAS 5-26555} except Mrk 231 and 
NGC 6240. The image for Mrk 231 is from deep observations made by the 
Schmidt telescope at the Xinglong Station of NAOs of China (Wu, Deng \& Xia 
2001), and the image for NGC 6240 is taken from the Digital Sky Survey. 
The soft X-ray contours are at 3, 5, 10, 20, 30, 50, 100 and 300$\sigma$ 
significance levels, respectively. It is obvious from Fig. Fig. 1 
that the soft X-ray emissions enclose the main optical images. In several 
cases (e.g., NGC 6240, Mrk 273, Mrk 231 and Arp 220), the soft X-ray 
emissions encompass the whole optical images including the long tidal tails. 
As {\it Chandra} has sub-arcsecond resolutions, its point-spread-function 
(PSF) is much smaller than the optical sizes, and hence the soft X-ray 
emissions are extended.
The most striking feature from these X-ray images is that for most objects
the central regions (of size $\sim 10\kpc$) are significantly brighter
than the outer diffuse emissions; sharp transitions appear to exist in the
diffuse X-ray emissions. These sharp transitions at $\sim$ 10 \kpc\ are
not related to central AGNs because the contributions from AGNs in the soft
X-ray band only dominate in the central $\sim$1$\arcsec$ region and it
decreases to less than 10\% of the source's surface brightness at
$\sim$3$\arcsec$ ($1\arcsec$ corresponds to $\sim$ 0.9\,\kpc\ at
redshift 0.04). Detailed fittings for Mrk 273 seem to support this.
Mrk 273 has a diffuse inner halo region ($\sim$10\kpc) and an outer halo
which extends to a few tens \kpc\ (Xia et al. 2002). Motivated by this, we
performed surface brightness profile analyses for the whole sample ULIRGs
in order to check whether such sharp transitions in the soft X-ray images
or breaks in the surface brightness profiles are common in ULIRGs.

Before we perform the surface brightness profile analyses, we need to
determine the center of the X-ray emissions. For most sample galaxies,
the dominant hard X-ray (2.0-8.0 keV) emission peak is selected because such
hard X-ray emission peaks usually coincide with the central dominant
nuclear positions. However, both NGC 6240 and Arp 299 have two bright hard
X-ray emission peaks. For NGC 6240, we select the southern peak as the 
center given that the separation of two hard X-ray emission peaks is only 
1.4$\arcsec$ and the southern peak is 3 times brighter than the northern 
one (Komossa et al. 2003). For Arp 299, the separation of the two hard 
X-ray emission peaks is $\sim$ 22$\arcsec$ and their intensities are similar 
(Ballo et al. 2004), hence we select the middle of the two peaks as the 
system center which also appears to be the geometric center of the outer 
X-ray contours. When we extract the surface brightness profiles of Arp 299, 
we remove the two nucleus regions. The  soft (0.3-2.0 keV) and hard (2.0-8.0 
keV) band radial surface brightness profiles were summed from the 
azimuthally-averaged and exposure-corrected narrower energy band radial 
surface brightness profiles extracted by the CIAO task {\it dmextract\/} for 
all ULIRGs. In this process, we subtracted all the bright point sources and 
bright emission clumps. However, the background is not subtracted and is 
instead directly fitted. To estimate the PSFs, we use the {\it Chandra}
Ray Tracer (ChaRT) in  both the soft and hard X-ray bands for each of our
target ULIRGs. ChaRT is a web interface to the SAOsac raytrace code which
was developed by the {\it Chandra} X-ray Center (CXC) for calibration
purposes. ChaRT traces rays through the {\it Chandra} X-ray optics (High
Resolution Mirror Assembly) to produce PSFs. Since ChaRT runs the same code
used internally at the CXC for calibration, it yields the best available
PSF for a point source at any off-axis angle and for any energy spectrum.

Fig. 2 shows the surface brightness profiles as well as
the PSFs in both the soft and hard X-ray bands for all sample ULIRGs;
the PSFs are normalized to the peaks in the surface brightness
profiles. We can see from Fig. 2 that all ULIRGs have
diffuse soft X-ray emissions. Moreover, extended hard X-ray emissions on
scales about 10 \kpc\ are also detected for some of ULIRGs such as NGC
6240 and IRAS 17208-0014 (cf. Franceschini et al. 2003). Furthermore, a break
in the surface brightness profiles appears to be common. The significance
is higher for some ULIRGs, notably, Mrk 273, NGC 6240, Mrk 231 and Arp 220,
which have very extended hot gaseous halos. We use the $\beta$-model, 
described by
\begin{equation}
S(r)=S_0\left[1+\left(\frac{r}{r_{\rm c}}\right)^2\right]^{-3\beta+0.5},
\end{equation}
to fit the surface brightness profiles for all sample galaxies; here
$r_{\rm c}$ is a core radius, $\beta$ is a power-law index and $S_0$ is an
overall normalization. The background is modeled as a constant. A double 
$\beta$-model gives a better description for the surface brightness profiles 
of Mrk 273, Mrk 231, NGC 6240 and Arp 220 than a single $\beta$-model. 
Specifically, a single $\beta$-model with three parameters ($S_{\rm 0}, 
r_{\rm c}$ and $\beta$) for Mrk 273, Mrk 231, NGC 6240 and Arp 220 yields 
a $\chi^2$ of 112.4, 82.0, 411.5 and 110.4 for 56, 85, 146 and 84 degrees 
of freedom, corresponding to a reduced $\chi^2$ of 2.01, 0.96, 2.82 and 1.31, 
respectively. With two (For Mrk 273, Mrk 231 and NGC 6240, $\beta$ is fixed 
in the second $\beta$-model) or one (For Arp 220, both of the two $\beta$ 
are fixed) additional parameter(s) in the double $\beta$-model, the $\chi^2$ 
improves to 72.4, 73.5, 197.6 and 102.2, corresponding to the reduced 
$\chi^2$ of 1.34, 0.89, 1.37 and 1.23, respectively. The standard $F$-test 
(e.g., Lupton 1993) shows that the confidence levels that the two or one 
additional parameters are not significant are $1.8\times 10^{-5}$, 
$4.3 \times 10^{-2}$, $2.7\times 10^{-23}$ and $2.3\times 10^{-2}$, 
respectively, for these four galaxies. Clearly a double-$\beta$ model 
improves the fits to Mrk 273 and NGC 6240 very significantly, while for the 
other two cases, the significance levels are more modest (at $\sim 2\sigma$ 
levels).  For the remaining 6 ULIRGs, a single $\beta$-model can fit the 
surface brightness profiles well. Also Note that for Mrk 231, Arp 220 and 
IRAS 05189-2524, the inner-most one or two points are not included in our 
fitting because these inner-most points are significantly affected by the 
central AGNs. From Fig. 2, we can also determine the break 
points in the surface brightness profiles. For objects fitted by double 
$\beta$-models, the break point is decided from the crossing point of the 
two $\beta$-models. On the other hand, for the six objects fitted with a 
single $\beta$-model, we take the first point that clearly deviates from the 
$\beta$-model as the break point. The break radius in the surface brightness 
profile is indicated by a vertical line in Fig. 2. As can be 
seen the break points decided from the surface brightness fits roughly agree 
with visual impressions. The break in the surface brightness profiles may
indicate different origins for the hot gas inside and outside the break
radius. Hence we investigate these two diffuse hot gaseous halos separately.
For convenience, hereafter we refer the inner X-ray diffuse gas as the
{\it inner halo} and  the more extended outer diffuse gas as the {\it outer
halo}. Table 2 presents the surface brightness fitting results, $r_{\rm c}$
and $\beta$, the physical scales for the inner and outer halos, as well as 
the merging state for each ULIRG, as indicated by the number of nuclei. We can 
see from Table 2 that there is no correlation between the extension of the 
soft X-ray halo and the merging state. The very extended hot gaseous halos 
have been detected for ULIRGs with both single nucleus and double nuclei. 
Notice that the physical scales for the inner halos of all sample ULIRGs are 
about 10\,kpc. We will investigate the spectral properties for the inner halos 
and outer halos separately in the following section.

\section{SPECTRAL ANALYSIS}

The sub-arcsecond resolution of {\it Chandra} allows us to perform reliable
spatially-resolved spectral analyses for the diffuse hot gaseous halos in
ULIRGs. We extracted the 0.3-8.0 keV spectra for the inner halos, but for the
outer halos, we mainly analyze the spectra in the 0.3-2.0 keV energy range
because there are few 2.0-8.0 keV photon counts in the outer halos.
All our spectral analyses were performed using the software packages
XSPEC V11.20 (Arnaud 1996) and FTOOLS 5.2
(Blackburn 1995)\footnote{http://heasarc.gsfc.nasa.gov/ftools/}.
Background spectra are extracted from source-free regions adjacent to the
sample ULIRGs on the same ACIS chips. We have also corrected the degradation
of the low energy quantum efficiency for the contamination buildup on the
optical blocking filter, which is particularly pronounced at energies below
1 \kev. Before any model fitting, we first rebin the data such that each bin
contains a minimum of 15 counts.  All the uncertainties are quoted
at the 90\% confidence level unless otherwise mentioned.

\subsection{THE OUTER X-RAY HALOS}

 From Fig. Fig. 2, the hard X-ray emission of IRAS 05189-2524
coincides with the  hard X-ray PSF, so the hard X-ray emission is
dominated by the central point source. Furthermore, there is no soft
X-ray outer halo for this galaxy. For UGC 05101 and IRAS 23128-5919,
their spectra can not be fitted well by either a MEKAL model (Mewe,
Gronenschild \& van den Oord 1985; Liedahl, Osterheld \& Goldstein 1995)
or a power-law model, so we do not include them in our subsequent analyses
for the outer halo. Below we present the spectral analysis results for
the outer halos for the remaining 7 ULIRGs. Fig. 3 shows
the spectra of the outer halos of these 7  ULIRGs. As can be seen, a MEKAL
thermal model gives an acceptable fit to all the galaxies. Table 3 lists
all the fitting parameters.

It is clear from Table 3 that the temperatures of the outer halos are 
about 0.6\keV\ and the metallicity is $\sim 0.1Z_\odot$ for NGC 6240, Mrk 
273, Mrk 231, Arp 220, Arp 299, IRAS 17208 and IRAS 20551, similar to those 
obtained using {\it Chandra} observations for Mrk 273 (Xia et al. 2002). 
NGC 6240 has the largest number of counts (2596 counts) detected within the 
outer halo, so the results for this galaxy are most reliable. In order to 
obtain more convincing results for other galaxies that have fewer counts, 
we first stack the data for Mrk 273, Mrk 231 and Arp 220 and then perform a 
spectral analysis on the stacked data. Arp 299 is not included because it 
was observed by ACIS-I detector instead of ACIS-S; we also do not include 
IRAS 17208-0014 and IRAS 20551-4250 because their photon numbers are too few 
to have any significant impact on the final results.

The stacking analysis is performed using FTOOLS task {\it addspec}, which adds
PHA spectra and background PHA files; detector redistribution and ancillary
response datasets are also combined appropriately, and the PHA channels of the
spectra files are remapped to correct for the effects of different redshifts.
The spectrum for the stacked data of Mrk 273, Mrk 231 and Arp 220 is shown
in the left panel of Fig. 4 and the fitting parameters are
shown in Table 3. It is reassuring that the temperature and metallicity
of the outer halo for NGC 6240 and the stacked data for Mrk 273, Mrk 231, Arp
220 (with more than 1200 counts) yield similar results: the temperature
is about 0.6 \keV\ and the metallicity is $0.1 Z_\odot$.

As a further check, we performed another test by fitting the data for NGC
6240, Mrk 273, Mrk 231 and Arp 220 simultaneously. In this analysis, we
fit each spectrum with a MEKAL model. However, we assume the metallicity
parameter is the same leaving just the temperature and normalization
parameter free for each object. The right panel in Fig. 4
shows the simultaneous fitting results for NGC 6240, Mrk 273, Mrk 231 and
Arp 220. The metallicity is about $0.1 Z_\odot$.

The above stacking analysis and simultaneous fitting for these ULIRGs
with more than several hundred detected counts demonstrate that
the outer halo of ULIRGs is from low metallicity hot gas. Reassuringly,
the low-metallicity for the outer halo of NGC 6240 is also confirmed by
the {\it XMM-Newton} data  (Y. J. Xue 2003, private communication). The
existence of a very extended hot gaseous halo with low metallicity
indicates that the outer hot gaseous halo has not yet been contaminated by
starbursts and/or AGN when the merging has almost finished. We return to
the origin of the low-metallicity outer halos in the discussion.

\subsection{THE INNER X-RAY HALOS}

Previous analyses for ULIRGs based on {\it Chandra} and {\it XMM-Newton}
data indicate that X-ray emissions of ULIRGs in the inner part
include contributions from both diffuse hot gas and X-ray binaries. For
some ULIRGs, the central obscured AGNs also make substantial contributions.
One key piece of evidence for AGNs comes from the detection of neutral
$\Feka$ lines (at 6.4\,keV), which are thought to arise from the iron
fluorescent emission by cold molecular materials illuminated by AGNs.
$\Feka$ lines are detected in roughly one half of the ULIRGs observed by
{\it Chandra} or {\it XMM-Newton} (for NGC 6240, Mrk 273, IRAS 19254-7245,
Mrk 231 and Arp 299, the $\Feka$ lines are certainly detected; for UGC 05101,
IRAS 20551-4250 and IRAS 23128-5919, $\Feka$ lines are marginally detected).
So unlike the extended soft X-ray outer halos,  which appear to be
dominated by the thermal emission of hot diffuse gas, the X-ray emissions
in the inner halos of a ULIRG can have several components such as the diffuse 
hot gas, X-ray binaries and central AGNs. As a result, the spectral analysis 
for the inner halo is much complicated.

Fig. 5 shows the (complex) spectra of the inner halos.
For Mrk 273, Mrk 231, Arp 220 and IRAS 05189-2524, the spectra
have three components, namely, a heavily absorbed power-law (for Mrk 273,
plus a narrow 6.4 keV $\Feka$ line), a less absorbed power-law and a MEKAL
thermal plasma emission plus line features (in particular Fe L and Ne line
complex, and other emission lines from $\alpha$-elements, such as $\Mgka$,
$\Sika$ lines, OVII and OVIII lines). For NGC 6240, the spectrum is well
fitted by two MEKAL thermal plasma emission components plus line features,
an absorbed power-law plus $\Feka$ lines. For the remaining ULIRGs,
two-component fits are acceptable because the detected counts are
insufficient to perform more detailed spectral analyses.

Given that our main goal is to study the properties and the origin of
diffuse hot gas instead of the central AGNs, in Table 4 we only list the
spectral parameters for the MEKAL thermal plasma emission component.
For NGC 6240, we only list the parameters for the stronger component of
the two MEKAL thermal emissions. From Table 4, the temperatures of the
inner halos are in the range of 0.6-0.8 \keV, with the exception of Mrk
231 which has a slightly lower temperature (0.4 \keV). Compared with the
outer halos, the metallicities of the inner halos are much higher,
$\ga 1 Z_\odot$. The high metallicity is likely due to contaminations by
stellar processes, such as stellar winds or superwinds that occur in massive
starburst galaxies; we discuss this in more details below.

\section{SUMMARY AND DISCUSSION}

Our analyses suggest that there are two distinct regions of hot diffuse
gas for a large fraction of ULIRGs. In particular, all ULIRGs have soft
X-ray emissions within $\sim$ 10\,kpc with $Z \sim 1 Z_\odot$ and
$kT \sim 0.7$\,keV. More extended ($\ga 10$\,kpc) hot gaseous halos exist for
some ULIRGs, but they have much lower metallicities ($Z \sim 0.1 Z_\odot$).
These results raise interesting questions about both the origin of hot gas
in elliptical galaxies and the feedback and chemical enrichment processes.
We discuss these two issues in turn.

\subsection{THE ORIGIN OF HOT GASEOUS HALOS IN ELLIPTICALS}

In our analysis, the separation between the inner and outer halos is
found from spatial analysis of the X-ray emission (see \S3), the
separation between these two components is reasonable but not clear-cut.
However, there is independent support for this separation. For example, 
Colina, Arribas \& Clements (2004) performed a careful study of the 
two-dimensional kinematics and studied ionization properties of the warm 
ionized gas in  Arp 220 based on data obtained from integral field optical 
spectroscopy. They identified two dynamical distinct regions from the surface 
brightness profiles and geometries, namely plumes and lobes that correspond 
to the inner halo and outer halo, respectively.

Fig. 6 shows the $\Lx$-T relation for both the outer
(top panel) and inner (bottom panel) halos of ULIRGs based on Table 5.
For comparison, we also plot the data for elliptical galaxies from
O'Sullivan et al. (2003a) and groups of galaxies from Xue \& Wu (2000).
 From Fig. 6, both the outer and inner halos of the
hot diffuse gas of ULIRGs overlap with the elliptical galaxies or groups
of galaxies. As the 39 elliptical galaxies in O'Sullivan et al.'s sample
are all luminous early-type galaxies, a straightforward interpretation
is that the hot gaseous halos of mergers resemble those of elliptical galaxies
and groups of galaxies, and can be regarded as another piece of evidence
for elliptical galaxy formation via mergers. However, notice that the
X-ray luminosities span two orders of magnitude for a given temperature.
Furthermore, all the measurements for elliptical galaxies and groups of
galaxies used in Fig. 6 are based on {\it ROSAT} or {\it
ASCA} data. Unlike {\it Chandra}, they cannot spatially resolve different
components of X-ray emission, and so the comparison between our data and
those of ellipticals and groups of galaxies is not direct. Ideally, one
should collect a sample of elliptical galaxies observed by {\it Chandra}
and perform the same analysis procedure and then compare the results; we
plan to address this issue in future work.

From Table 5, we can also see that the contributions of the inner and outer 
halos to the total luminosity of the diffuse hot gas vary for different 
ULIRGs. For most ULIRGs, the inner halo dominates the luminosity of diffuse 
hot gas, while for some ULIRGs, such as Mrk 273 and Arp 220, the outer halo 
contributes more (about 2/3) to the total luminosity. As the outer halo has 
low metallicity ($\sim 0.1Z_\odot$) while the inner halo has high metallicity 
($\sim 1Z_\odot$), it is not difficult to understand the wide range of 
metallicities reported for the sample of ellipticals in O'Sullivan et al 
(2003a). If the inner halo is the main contributor to the X-ray luminosity, 
then their analysis will give a high metallicity for the total gas, and 
vice versa.

The high metallicity of the inner halos is not surprising, as they can be
easily produced by processes such as stellar winds and superwinds that are
expected to carry a substantial amount of heavy metals (Heckman 2001; see
\S5.2). The more surprising result is the low-metallicity of the outer
halos, which seems to be lower than those reported for groups of galaxies
(e.g., Buote et al. 2003; O'Sullivan et al. 2003b). However, the 
low-metallicity is not unique. For example, Sambruna et al. (2004) analysed 
the diffuse hot gaseous halo in the giant elliptical NGC 6251, a host
galaxy of a radio galaxy, based on {\it XMM-Newton} observations. They
find that the diffuse hot gaseous halo can be fitted by a thermal plasma model
plus a hard component. The thermal component has a temperature of 1.6 keV
and a low metallicity ($\sim 0.1Z_\odot$) while the hard component has a
very flat power-law with a photon-index of zero. {\it XMM-Newton} data for
NGC6240 are also consistent with a thermal component with $Z\sim
0.1Z_\odot$ plus a hard power-law component (Y.J. Xue 2003, private
communication). Our modeling so far does not include a hard component.
To test the sensitivity of our results to this component, we performed
spectral analyses for the outer halo of NGC 6240 and the stacked spectrum
for Mrk 273, Mrk 231 and Arp 220 in the 0.3-8.0 keV energy range. The results
are almost the same as those for NGC 6251 except the temperature is slightly
lower (0.6\,keV) if we use a MEKAL model plus a power-law hard component with 
the photon index $\Gamma$ fixed in a range from zero to $\sim 1.6$. Only
when a larger power-law index ($\Gamma\approx 2.5$) is adopted for the 
power-law component, the metallicity approaches $Z\sim 0.4Z_\odot$,  which 
is still sub-solar. The low-metallicity of the gas implies that the cooling 
time of the gas may have been under-estimated by a factor of few (see, e.g.,
Fig. 9-9 in Binney \& Tremaine 1987 for the dependence of the cooling rate
on metallicity) since many previous studies assume solar-metallicities
(e.g., Sarazin 1990). This hot gas may  remain X-ray luminous even after
the stellar components at the center have relaxed dynamically.

The low metallicity of the outer halos indicates that they are from
low-metallicity gas that has not yet been chemically contaminated. Some
ULIRGs likely involve the merging of two gas-rich spirals. The cold gas
in the halos of spirals is more metal-poor, and can have a metallicity
as low as $Z \sim 0.1 Z_\odot$. For example, HII regions located in the
disk-halo interface of NGC 55 has $Z \sim 0.1 Z_\odot$ (T\"ullmann et al.
2003); high velocity clouds in the halos of local group galaxies also
have such low metallicities (Wakker 2003). During the violent merging,
the metal-rich gas in the inner part will be driven toward the center as
they lose angular momentum (Barnes \& Hernquist 1996), while the outer
gas may be launched as tidal tails, which is then shock-heated when they
fall back. Such tidal tails are detected by HI observations, which can
contain up to half of the HI gas in progenitor galaxies (Hibbard \& Mihos
1995). The low metallicity may also be due to secondary infalls of
low-metallicity gas in group environment or continuous accretion from
ambient cosmological flow (e.g., Brighenti \& Mathews 1998, 1999; Mathews
\& Brighenti 2003). In all cases, we require that the outer gas has not
been contaminated by the feedback processes occurring at the center (for
more, see the next subsection).

Regardless of the origin of the hot gas in the inner parts and outer
parts, it is puzzling that some ULIRGs (such as Mrk 273, NGC 6240,
Mrk 231) have very extended hot gaseous halos while others have little
or no extended halos.
The bolometric luminosities of hot gaseous halos of NGC 6240 and Mrk 231
are in the order of $10^{42}\ergs$, which fall in the range of bolometric
luminosities for fossil groups of galaxies (Jones et al. 2003). We
estimate the hot gas mass of the outer halos for NGC 6240, Mrk 273, Mrk 231
and Arp 220, which are 2.3, 1.1, 1.3 and 0.1 times $10^{10}M_\odot$.
Note that these numbers are model-dependent and somewhat uncertain due to
limited photon numbers. The hot gas mass of outer halos for NGC 6240, Mrk
273, Mrk 231 and Arp 220 is in the range of elliptical galaxies with bright
X-ray emission that is $10^{9}M_\odot$ to $10^{10}M_\odot$ (Matsushita
2001). For comparison, the hot gas mass for the four groups of galaxies
observed by {\it XMM-Newton} ranges from $ 8\times 10^{9}M_\odot$ to
$4\times 10^{11}M_\odot$ (Figueroa-Feliciano et al. 2003). The hot gas
mass for the outer halos in NGC 6240, Mrk 273, Mrk 231 is comparable to
that in groups of galaxies. The cold gas mass of spiral galaxies is
typically $10^{9}M_\odot$ to $10^{10}M_\odot$ (Sansom et al. 2000) and the 
cold gas mass of Arp 220 and Arp 299 by HI detection is $2\times 
10^{9}M_\odot$ (Yun et al. 1999). The mass budget in the outer hot gaseous 
halos of NGC 6240, Mrk 273, Mrk 231 may be too large to be only from merging 
progenitor galaxies. Furthermore, from a VLA imaging survey of a distance 
limited sample of 9 ULIRGs, Yun et al (1999) reveals the presence of $\sim 
100\kpc$ scale giant radio plumes in only 3 objects: Mrk 231, Mrk 273 and 
NGC 6240. Intriguingly, these three objects have the most extended low 
metallicity hot gaseous halos in our sample. Moreover, among the ULIRGs 
observed by {\it XMM-Newton} (Franceschini et al. 2003), Mrk 231 and IRAS 
19254-7245 have very extended hot gaseous halos, while IRAS 19254-7245 has 
excess radio emission detected by the Parkes-MIT-NRAO survey with a radio 
luminosity $\L6.3\approx 3.5\times 10^{39}\,\ergs$ (Roy \& Norris 1997). The 
radio emissions may be produced by relativistic electrons accelerated in 
shocks in these violent merging galaxies. Given that powerful radio sources 
inhabit rich environments at high redshift but are typically found in groups 
or poor clusters at low redshift (Best et al. 2002), it is plausible that 
low redshift powerful radio sources such as Mrk 231, Mrk 273 and NGC 6240 
were in group environments before merging.

Given that the ULIRGs have extended hot gaseous halos and those without both
exhibit single and double nuclei morphologies, the merging state cannot be
the main origin for the difference. It is possible that the progenitors of
our sample galaxies are different: the ULIRGs with very extended
hot gaseous halos may be evolved from group of galaxies and others from gas
rich spirals and the extended hot gas of such systems is from HI tidal tail.
The group origin is also corroborated by the kinematic data. Table 2 lists
the velocity dispersions of ULIRGs available from Tacconi et al. (2002).
Mrk 273 and NGC 6240 have high velocity dispersion ($\sim 300\kms$), similar
to groups of galaxies. If these velocity dispersions reflect the true
potential well, then the kinematic data support the group origin for Mrk
273 and NGC 6240. For Mrk 231, the measured velocity dispersion is low
($\sim \,120\kms$), but this low value may be because its AGN hot dust
continuum dilutes the stellar continuum and reduces the absorption features
and hence the velocity dispersion; Tacconi et al. (2002) argued that this
reduction is not significant. The group connection for Mrk 231 is therefore
less clear.

\subsection{CONSTRAINTS ON FEEDBACK AND CHEMICAL ENRICHMENT}

 From Table 2, we see that the radius for the metal-rich inner halos for
all ULIRGs is about $\sim 10 \kpc$, regardless of whether the merging
is still ongoing or has largely completed (as indicated by the number of
nuclei). Furthermore, {\it Chandra} observations with HRC or ACIS reveal
many features, such as filaments, arcs and loops, for these inner X-ray
halos. These X-ray features show close spatial correlations with
$\Halpha$ emissions, for example, in NGC 6240 (see Fig. 2 in Lira et al. 
2002), Mrk 273 (see the left panel of Fig. 6 of Xia et al. 2002) and Arp 
220 (see Fig. 9 in McDowell et al. 2003). Such correlations have
also been found for nearby starburst galaxies M82, NGC 253 and other
edge-on star forming galaxies. The popular explanation for the correlation
between X-ray emissions and optical line emissions is that a pre-existing
halo medium is heated by a starburst driven superwind or by metal-enriched
wind material compressed in reverse shocks near the walls of the outflow
(Strickland et al. 2003).
The physical scale for these nearby star forming galaxies is 4.5-9 \kpc,
similar to the scale of inner halos in ULIRGs.
There has also been a reported over-abundance of Si and S compared with
other $\alpha$-elements in M82 (Umeda et al. 2002). The abundances of Si
and S inferred from our X-ray spectra are rather uncertain, but consistent
with the pattern seen in M82.

The star formation rates in ULIRGs, starburst and star-forming galaxies
differ by two orders of magnitude. But it appears that they show similar
correlations between X-ray emissions and H$\alpha$ emissions, their
morphology, physical scales and the (rather uncertain) S and Si
over-abundance are all similar. These similarities suggest that feedback
processes are not determined by the star formation rate alone but also
by other physical parameters (see below). Colina et al (2004) performed
a study of the two-dimensional kinematics of Arp 220. They find that the
inner halo (plumes) of Arp 220 is under-luminous in both the soft X-ray
and $\Halpha$ emissions at least by a factor of 10 with respect to the
starburst galaxies M82 and NGC 2992 if one normalizes these systems by
their infared luminosities which are a measure of their star formation
rates. This suggests that the starburst-driven wind in Arp 220 is much
less efficient in producing the X-ray and $\Halpha$ emissions than starburst
galaxies and normal spirals. Their explanation is that the ambient
interstellar medium in ULIRGs is much denser, and hence the mass loading
in the superwinds will be greater. As a result, galactic winds will be
slowed down faster and terminate at smaller radii. Also the planar geometry
of  galactic disks may allow winds to escape more easily perpendicular to
the planes; in ULIRGs the ambient medium may be quasi-spherical due to
violent merging, so wind cannot readily escape. As a result, the chemical
enrichment due to the superwinds in ULIRGs cannot proceed to very large
radii. This effect is also seen in the numerical simulations of multiple
galaxy mergers by Bekki (2001) who found that metals produced and ejected
in the central star formation regions are mostly mixed with the local
interstellar media. The feedback and chemical enrichment processes are
hence, not surprisingly, determined not only by the overall star formation
rate but also by local interstellar medium properties.

\acknowledgments

We would like to thank the {\it Chandra} X-ray Center (CXC) and the NASA
High-Energy Astrophysics Science and Research Center (HEASARC). We also
thank Drs. Y. Gao, X. P. Wu, J. Hibbard and S. White for advice and helpful
discussions. Thanks are also due to C. N. Hao for help in the optical data
reduction. This project is supported by the NSF of China No. 10333060 and 
TG1999075404. X. Y. Xia and Z. G. Deng gratefully acknowledge the exchange 
program between NSFC and DFG. X. Y. Xia also thanks the financial support 
by the visitor program at Jodrell Bank Observatory. S. Mao wishes to thank 
the hospitalities of Shanghai Astronomical Observatory in particular
Dr. C. G. Shu during a visit in 2003.

\clearpage


\figcaption[fig1.ps]{\label{fig:smoothed}
The left panels show the adaptively smoothed and exposure-corrected 
soft X-ray (0.3--2.0 keV) images for 10 sample ULIRGs, with a maximum 
smoothing scale of 10 pixels ($\approx 5\arcsec$). The right panels 
show the optical images overlaid by soft X-ray contours at 3, 5, 10, 
20, 30, 50,100 and 300$\sigma$ levels, respectively.
}

\figcaption[fig2.ps]{\label{fig:surface}
Soft (0.3--2.0 keV, filled circles) and hard X-ray (2.0--8.0 keV, open
triangles) azimuthally-averaged and exposure-corrected surface brightness 
profiles for 10 sample ULIRGs, respectively. The solid line presents the 
best-fit model, the dotted line presents the $\beta$-model(s) and the 
approximate soft X-ray background level is indicated by a thick dashed 
line. The PSFs of all objects except Arp 299 for both the soft (short-dashed 
line) and hard X-ray (long-dashed line) are also shown for comparison. For 
Arp 299, the center for surface brightness profile analysis is not the peak 
of hard X-ray emission (see \S3), so no PSFs are plotted. The vertical 
dot-dashed line in each panel indicates the break point in the surface 
brightness profile.
}

\figcaption[fig3.ps]{\label{fig:spectra1}
The 0.3--2.0 keV spectra for the outer halos of 7 ULIRGs (see \S2). A MEKAL
model with temperature of about 0.6\,keV and metallicity of about
$0.1Z_\odot$ provides a satisfactory fit for most objects. The fitting
parameters are given in Table 3.
}

\figcaption[fig4.ps]{ \label{fig:spectra2}
The left panel is the stacked spectrum for Mrk 273, Mrk 231 and Arp 220;
the right panel from bottom to top shows the simultaneous fitting to
the spectra for Arp 220, Mrk 231, Mrk 273 and NGC 6240, respectively 
(see \S3).
}

\figcaption[fig5.ps]{ \label{fig:spectra3}
The 0.3--8.0 keV spectra for the inner halos for all sample ULIRGs. Each
spectrum is fitted well with the superposition of heavily (or less) absorbed
power-law model(s), a thermal MEKAL component(s) plus $\Feka$ lines and
other $\alpha$-element emission lines. The fitting parameters are given
in Table 4.
}

\figcaption[fig6.ps]{\label{fig:Lx-T}
The top and bottom panels show the $\Lx$-T relation for the outer halos
and inner halos of ULIRGs (filled circles), respectively. For comparison,
we also plot data for elliptical galaxies (crosses) from O'Sullivan et al.
(2003a) and groups of galaxies (triangles) from Xue \& Wu (2000).
}

\clearpage

\begin{deluxetable}{lccccl}
\tablewidth{430pt}
\tablecaption{Chandra Observations Log}
\tablehead{
\multicolumn{1}{l}{Source} & \multicolumn{1}{c}{Redshift} &
\multicolumn{1}{c}{Obs. ID} &
\multicolumn{1}{c}{Exposure (ks)\tablenotemark{a}} & \multicolumn{1}{c}{Detector} &
\multicolumn{1}{l}{Obs. Mode} }

\startdata
Mrk 273  & 0.0378 & 809  & 40.0 & ACIS-S & VFAINT \\
Mrk 231  & 0.0422 & 1031 & 37.8 & ACIS-S & FAINT  \\
NGC 6240 & 0.0245 & 1590 & 36.7 & ACIS-S & FAINT  \\
Arp 220  & 0.0181 & 869  & 55.2 & ACIS-S & VFAINT \\
Arp 299  & 0.0103 & 1641 & 24.7 & ACIS-I & FAINT  \\
IRAS 05189-2524\tablenotemark{b} & 0.0426 & 2034/3432 & 6.0/14.9 & ACIS-S & FAINT \\
IRAS 17208-0014 & 0.0428 & 2035      & 47.6     & ACIS-S & FAINT \\
IRAS 20551-4250 & 0.0428 & 2036      & 37.6     & ACIS-S & FAINT \\
IRAS 23128-5919 & 0.0446 & 2037      & 32.8     & ACIS-S & FAINT \\
UGC 05101       & 0.0394 & 2033      & 47.9     & ACIS-S & FAINT \\
\tablenotetext{a} {The background flare time period has been excluded.}
\tablenotetext{b} {Only observation ID 3432 are used in this paper.}
\enddata
\end{deluxetable}

\begin{deluxetable}{lllllccl}
\tablecaption{The merger stages, results of surface brightness fits
 and scales of hot gaseous halos}

\tablehead{
\colhead{Source} & \colhead{N of nuclei} & \colhead{d\tablenotemark{a}} &
\colhead{$r_c$\tablenotemark{b}} & \colhead{$\beta$} &
\colhead{$R_1$\tablenotemark{c}} & \colhead{$R_2$\tablenotemark{d}} &
\colhead{$\sigma$\tablenotemark{e}}}

\startdata
Mrk 273   & double & 0.81 & $2.75\pm0.06$  & $0.83\pm0.02$ & 10.5 & $43.6\times23.6$ & 281 \\
          &        &      & $34.15\pm2.07$ & 1.0 (fixed)   &      &                  &     \\
Mrk 231   & single &      & $3.70\pm0.21$  & $0.65\pm0.03$ &  9.9 &        36.8      & 120 \\
          &        &      & $30.40\pm1.80$ & 1.0 (fixed)   &      &                  &     \\
NGC 6240  & double & 0.58 & $3.73\pm0.04$  & $1.01\pm0.01$ &  9.3 & $42.0\times34.0$ & 324 \\
          &        &      & $28.52\pm0.65$ & 1.0 (fixed)   &      &                  &     \\
Arp 220   & double & 0.35 & $4.44\pm0.25$ & 1.0 (fixed)   &  5.0 & $19.8\times10.3$ & 166 \\
          &        &      & $14.83\pm0.84$ & 1.0 (fixed)   &      &                  &     \\
Arp 299   & double & 4.40 & $7.74\pm0.21$ & 1.0 (fixed)   & 12.4 &        13.6      &     \\
IR 05189  & single &      & $4.29\pm0.39$  & 1.0 (fixed)   &      &        11.8      &     \\
IR 17208  & single &      & $3.79\pm0.17$  & $0.88\pm0.04$ &  8.0 &        15.5      & 229 \\
IR 20551  & double & 0.92 & $2.38\pm0.08$  & $0.71\pm0.02$ & 15.0 &        17.3      & 141 \\
IR 23128  & double & 4.37 & $3.84\pm0.14$  & $0.91\pm0.03$ & 10.9 &        14.2      &     \\
UGC 05101 & single &      & $0.90\pm0.05$  & $0.60\pm0.01$ &  8.7 &        14.3      &     \\

\tablenotetext{a} {$\rm d$ is the projected distance between two nuclei
in units of kpc. The projected distance for Mrk 273, NGC 6240 and Arp 220
is from Scoville et al. (2000), IR 20551 and IR 23128 from Cui et al.
(2001) and Arp 299 from Zezas et al. (2003).}
\tablenotetext{b} {$r_c$ is the core radius of the $\beta$-model in units of kpc.}
\tablenotetext{c} {$R_1$ is the radius of the inner halo in units of kpc.}
\tablenotetext{d} {$R_2$ is the scale of the outer halo in units of kpc.
For a circular shape we quote the radius only; for an elliptical shape,
both major and minor radii are given.}
\tablenotetext{e} {velocity dispersions in unit of $\kms$, taken
from Tacconi et al. (2002).}
\enddata
\end{deluxetable}

\begin{deluxetable}{lccccccc}
\tablewidth{500pt}
\tablecaption{Spectral parameters for the outer halos}
\tablehead{
\colhead{Source} & \colhead{Counts} & \colhead{Galactic $\NH$\tablenotemark{a} } & \colhead{kT} & \colhead{Z} &
\colhead{$\chi^{2}$/d.o.f} & \colhead{$A_{\rm MEKAL}$\tablenotemark{b}}
& \colhead{$\Lx$\tablenotemark{c}} \\
&   & \colhead{[$10^{20}\ {\rm cm}^{-2}]$} & \colhead{[keV]}
   & \colhead{[$Z_\odot$]}    & & \colhead{[$10^{-5}$]}
   & \colhead{[$10^{41}\ergs$]} }
\startdata
Mrk 273  & 520 & 1.1 & $0.54^{+0.06}_{-0.07}$ & $0.12^{+0.05}_{-0.04}$ & 31.3/29 & $7.76^{+1.84}_{-2.12}$ & $1.62^{+0.38}_{-0.42}$ \\
Mrk 231  & 357 & 1.3 & $0.63^{+0.08}_{-0.09}$ & $0.08^{+0.06}_{-0.03}$ & 25.2/22 & $6.84^{+2.08}_{-1.94}$ & $1.63^{+0.47}_{-0.43}$ \\
NGC 6240 & 2596  & 6.0 & $0.63^{+0.03}_{-0.04}$ & $0.09^{+0.03}_{-0.02}$ & 78.8/82 & $69.15^{+11.95}_{-11.05}$ & $5.88^{+1.12}_{-0.98}$ \\
Arp 220  & 503 & 4.3 & $0.62^{+0.07}_{-0.06}$ & $0.08^{+0.04}_{-0.03}$ & 24.7/30 & $7.36^{+1.74}_{-1.78}$ & $0.32^{+0.07}_{-0.08}$ \\
Arp 299  & 427 & 1.1 & $0.60^{+0.09}_{-0.09}$ & $0.12^{+0.21}_{-0.05}$ & 35.4/26 & $20.23^{+9.47}_{-10.93}$ & $0.34^{+0.16}_{-0.18}$ \\
IR 17208 & 60 & 10.1 & $0.64^{+0.35}_{-0.29}$ & $0.11^{+0.10}_{-0.08}$ & 4.7/7 & $1.23^{+0.50}_{-0.52}$ & $0.34^{+0.14}_{-0.14}$ \\
IR 20551 & 62 & 3.8 & $0.30^{+0.17}_{-0.07}$ & $0.16^{+0.11}_{-0.09}$ & 1.6/4 & $1.84^{+0.58}_{-0.61}$ & $0.42^{+0.13}_{-0.13}$ \\
Stacked\tablenotemark{d} & 1380 & & $0.58^{+0.05}_{-0.05}$ & $0.10^{+0.03}_{-0.03}$ & 46.1/63 & $6.66^{+1.19}_{-1.13}$ &  \\

\tablenotetext{a} {Galactic $N_H$ values are taken from Dickey \&
Lockman (1990) using the LHEASOFT tool nh (http://heasarc.gsfc.nasa.gov)}
\tablenotetext{b} {The model normalization, $A_{\rm MEKAL}$, is defined
as $10^{-14} \int n_{\rm e} n_{\rm H} dV / (4 \pi (1+z)^2D_A^2)$, where
$n_{\rm e}$ is the electron number density (\pcmcu), $n_{\rm H}$ is the
hydrogen number density (\pcmcu) and $D_A$ is the angular diameter
distance to the source in units of cm.}
\tablenotetext{c} {The luminosity in the 0.3-2.0\,keV band, corrected
for the Galactic but not intrinsic absorptions; the latter may be small
in the outer halos.}
\tablenotetext{d} {Fit parameters for the stacked spectrum for Mrk 273,
  Mrk 231 and Arp 220.}
\enddata
\end{deluxetable}

\begin{deluxetable}{lccccccc}
\tablewidth{510pt}
\tablecaption{Spectral parameters for the inner halos}
\tablehead{
\colhead{Source} & \colhead{Model} & \colhead{$N_H$ \tablenotemark{a}} & \colhead{kT} & \colhead{Z} &
\colhead{$\chi^{2}$/d.o.f} & \colhead{$A_{\rm MEKAL}$\tablenotemark{b}}
& \colhead{$\Lx$\tablenotemark{c}} \\
&   & \colhead{[$10^{21}\ {\rm cm}^{-2}]$} & \colhead{[keV]}
   & \colhead{[$Z_\odot$]}    & & \colhead{[$10^{-5}$]}
   & \colhead{[$10^{41}\ergs$]} }
\startdata
Mrk 273 &T+2PL &  & $0.78^{+0.07}_{-0.07}$ & $0.99^{+1.73}_{-0.21}$ & 98.1/88 & $
1.45^{+0.18}_{-0.17}$ & $1.32^{+0.18}_{-0.12}$ \\
Mrk 231 &T+2PL & $3.95^{+2.45}_{-2.45}$ & $0.38^{+0.19}_{-0.15}$ & $1.04^{+0.68}_{-0.44}$ & 96.3/97 & $
6.26^{+0.85}_{-0.85}$ & $7.09^{+1.01}_{-0.99}$ \\
NGC6240& 2T+PL & $8.99^{+0.61}_{-0.79}$ & $0.73^{+0.05}_{-0.04}$ & $0.94^{+0.71}_{-0.39}$ & 213.6/201 &
$172.09^{+22.91}_{-20.09}$ & $65.56^{+8.44}_{-5.56}$ \\
Arp 220 & T+2PL & $0.66^{+0.54}_{-0.41}$ & $0.78^{+0.09}_{-0.12}$ & $0.99^{+1.33}_{-0.30}$ & 46.5/53 & $
1.03^{+0.14}_{-0.15}$ & $0.22^{+0.03}_{-0.03}$ \\
Arp 299 & T+PL & $1.20^{+1.00}_{-1.00}$ & $0.65^{+0.05}_{-0.04}$ & $0.90^{+3.59}_{-0.27}$ & 194.6/133 &
$19.52^{+1.18}_{-1.22}$ & $1.31^{+0.09}_{-0.11}$ \\
IR 05189& T+2PL &  & $0.81^{+0.26}_{-0.21}$ & $1.08^{+5.38}_{-0.58}$ & 111.1/112 & $0.55
^{+0.27}_{-0.28}$ & $0.67^{+0.33}_{-0.34}$ \\
IR 17208& T+PL & $8.67^{+4.13}_{-8.67}$ & $0.73^{+0.78}_{-0.38}$ & $1.11^{+0.96}_{-0.70}$ & 23.6/19 & $3.44^{+0.74}_{-0.82}$ & $4.59^{+1.01}_{-1.09}$ \\
IR 20551 & T+PL & $0.19^{+0.35}_{-0.19}$ & $0.66^{+0.07}_{-0.15}$ & $0.97^{+0.55}_{-0.17}$ & 42.2/33 & $1
.69^{+0.19}_{-0.18}$ & $2.10^{+0.20}_{-0.20}$ \\
IR 23128 & T+PL &  & $0.56^{+0.10}_{-0.11}$ & $0.98^{+1.66}_{-0.32}$ & 46.9/40 & $1.24^{+0.19}_{
-0.18}$ &$1.70^{+0.30}_{-0.20}$  \\
UGC 05101 & T+PL & $3.37^{+5.63}_{-3.37}$ & $0.61^{+0.16}_{-0.44}$ & $1.14^{+0.67}_{-0.28}$ & 34.0/20 &
 $0.84^{+0.21}_{-0.21}$ & $1.04^{+0.26}_{-0.24}$ \\

\tablenotetext{a} {The intrinsic absorption of MEKAL model.}
\tablenotetext{b} {The model normalization, $A_{\rm MEKAL}$, is as defined
in Table 3.}
\tablenotetext{c} {The luminosity in the 0.3-10\,keV band, corrected
for both the Galactic  and intrinsic absorptions.}
\enddata
\end{deluxetable}

\begin{deluxetable}{lcccccc}
\tablecaption{The temperatures and bolometric luminosities of the inner and outer halos}
\tablehead{
\colhead{Source} & \colhead{kT(inner)} & \colhead{$\Lx$(inner)} &
\colhead{kT(outer)} & \colhead{$\Lx$(outer)}
& \colhead{Ratio\tablenotemark{a}} & \colhead{Mass\tablenotemark{b}}\\
&    \colhead{[keV]} & \colhead{[$10^{42}\ergs$]}    &\colhead{[keV]}
   &\colhead{[$10^{42}\ergs$]}  & &\colhead{[$10^{10}M_\odot$]}}
\startdata
Mrk 273 &  $0.78^{+0.09}_{-0.04}$ & $0.18^{+0.02}_{-0.02}$ &  $ 0.54^{+0.06}_{-0.07}$ & $0.28^{+0.07}_{-0.07}$ & 61\% &1.10\\
Mrk 231 &  $0.38^{+0.19}_{-0.15}$ & $1.00^{+0.14}_{-0.13}$ &  $ 0.63^{+0.08}_{-0.09}$ & $0.29^{+0.09}_{-0.08}$ & 22\% &1.25 \\
NGC6240&   $0.73^{+0.05}_{-0.04}$ & $8.82^{+1.12}_{-1.03}$ &  $ 0.63^{+0.03}_{-0.04}$ & $1.04^{+0.17}_{-0.18}$ & 11\% &2.30\\
Arp 220 &  $0.78^{+0.09}_{-0.12}$ & $0.03^{+0.004}_{-0.004}$&  $ 0.62^{+0.07}_{-0.06}$ & $0.06^{+0.01}_{-0.01}$ &67\% &0.12\\
Arp 299 &  $0.65^{+0.05}_{-0.04}$ & $0.18^{+0.01}_{-0.01}$ &  $ 0.60^{+0.09}_{-0.09}$ & $0.06^{+0.03}_{-0.03}$ & 25\% &\\
IR 05189&  $0.81^{+0.26}_{-0.21}$ & $0.09^{+0.05}_{-0.05}$ &                          &                        &      &\\
IR 17208&  $0.73^{+0.78}_{-0.38}$ & $0.61^{+0.13}_{-0.14}$ &  $ 0.64^{+0.35}_{-0.29}$ & $0.06^{+0.03}_{-0.03}$ &9\%   &\\
IR 20551 & $0.66^{+0.07}_{-0.15}$  & $0.27^{+0.03}_{-0.03}$ &  $ 0.30^{+0.17}_{-0.07}$ & $0.09^{+0.03}_{-0.03}$ &25\% &\\
IR 23128 & $0.56^{+0.10}_{-0.11}$ & $0.22^{+0.03}_{-0.03}$ &                          &                          &    &\\
UGC 05101& $0.61^{+0.16}_{-0.44}$ & $0.13^{+0.03}_{-0.03}$ &                          &                          &    &\\
\tablenotetext{a} {The ratio of the outer halo luminosity to the total
  luminosity of the whole diffuse gaseous halo.}
\tablenotetext{b} {The hot gas mass of the outer halos.}

\enddata
\end{deluxetable}









\enddocument